\documentclass[aps,prd,groupedaddress,showpacs,tighten,11pt]{revtex4}
\usepackage{graphicx}
\usepackage{bm}
\usepackage{amsmath,amssymb}
\usepackage{latexsym}

\begin{document}

\title{Next-to-leading term of the renormalized stress-energy tensor 
of the quantized massive scalar field in Schwarzschild spacetime. 
The back reaction.}

\author{Jerzy Matyjasek}
\email{jurek@kft.umcs.lublin.pl, matyjase@tytan.umcs.lublin.pl}
\affiliation{Institute of Physics, 
Maria Curie-Sk\l odowska University\\
pl. Marii Curie-Sk\l odowskiej 1, 
20-031 Lublin, Poland}

\author{Dariusz Tryniecki}
\affiliation{Institute of Theoretical Physics, 
Wroc\l aw University\\
pl. M. Borna 9, 
50-204 Wroc\l aw, Poland}

\date{\today}

\begin{abstract}
The next-to-leading term of the renormalized stress-energy tensor of
the quantized massive field with an arbitrary curvature coupling in the
spacetime of the Schwarzschild black hole is constructed. It is
achieved by functional differentiation of the DeWitt-Schwinger
effective action involving coincidence limit of the Hadamard-
Minakshisundaram-DeWitt-Seely coefficients $a_{3}$ and $a_{4}.$ The
back reaction of the quantized field upon the Schwarzschild black hole
is briefly discussed.
\end{abstract}





\maketitle

\section{\label{intro}Introduction}
If the Compton length, $\lambda_{c} =\hbar/mc,$ associated with a
quantized massive field is much smaller than a characteristic radius
of curvature, $L,$ (where the latter means, as usual, any length scale
of the background geometry) then the nonlocal contribution to the
renormalized effective action, $W_{R},$ can be neglected and its
series expansion in $m^{-2}$ can be constructed using DeWitt-Schwinger
method. Since in the renormalization prescription one has to absorb
the first three terms of the expansion into the classical action of
the quadratic gravity with the cosmological term, the lowest
nonvanishing term of the $W_{R}$ is to be constructed from the
(integrated) coincidence limit of the fourth Hadamard-
Minakshisundaram-DeWitt-Seely coefficient, $[a_{3}],$ whereas next to
leading term is constructed form $[a_{4}].$ Generally one has
\begin{equation}
W_{R}\,=\,\frac{1}{32\pi^{2}}\sum_{n=3}^{\infty}
\frac{(n-3)!}{(m^{2})^{n-2}}\int d^{4}x \sqrt{g} [a_{n}].
                           \label{Weff}
\end{equation}
For the technical details of this approach the reader is referred, for
example, to Refs.~\cite{Barvinsky:1985an,FZ3} and the references cited
therein.

It is a well known fact that the complexity of  $[a_{n}]$ increases
rapidly with $n$ making calculations of the coefficients for $n>2$ a
highly nontrivial task. It is expected therefore that the
applicability of the series (\ref{Weff}), truncated at some definite
$n,$ will be limited to the simplest geometries with symmetries. On
the other hand, however, as the coefficients depend on the background
geometry, and, possibly, on a ``potential" term, they can be used to
construct the renormalized stress-energy tensor, $T_{a}^{b},$ by
functional differentiation of $W_{R}$ with respect to the metric. Such
a tensor can be defined in a wide class of geometries, and, by
construction, it gives a unique opportunity to study the back reaction
on the metric in a self-consistent way. Of course, the results of such
calculations should be interpreted with care as the particle creation,
which is a nonlocal process, is ignored.

The coefficient $[a_{2}]$ has been calculated by DeWitt~\cite{Bryce1}
whereas $[a_{3}]$ by has been obtained by Sakai and
Gilkey~\cite{Sakai,MR53:4150}; the fifth coefficient, $[a_{4}],$ has
been calculated in Refs.~\cite{Avramidi:1989ik,Amsterdamski,Ven}. The
results for $[a_{4}]$ are rather hard to compare as there are various
simplification strategies that can be employed, and, unfortunately,
some of the results contain not only typographical errors. Moreover, a
compact or even tricky notation is of little help in situations when
the main task is to calculate the stress-energy tensor in a specific
spacetime. Therefore, in order to construct the approximation to the
renormalized stress-energy tensor we have independently calculated
$[a_{4}]$ for a massive scalar field with an arbitrary curvature coupling
satisfying the equation
\begin{equation}
\left(-\Box +\xi R + m^{2} \right)\phi = 0,
                            \label{covKG}
\end{equation}
where $\xi$ is the dimensionless parameter describing the curvature
coupling and $R$ is the curvature scalar, using fully covariant method
of DeWitt~\cite{Bryce1} and checked the calculations constructing
$[a_{4}]$ in the Riemann normal coordinates~\cite{Parker}. The thus
calculated coefficients have been compared among themselves and with
their known values in concrete geometries. For example, when
specialized to $n=4,$ the coefficient $[a_{4}]$ precisely reproduces
the coefficient obtained from Dowker's general formula for $[a_{n}]$
in the de Sitter spacetime ~\cite{MRBrown}
\begin{eqnarray}
[a_{4}]^{dS} &=& -\frac{6}{(4a^{2})^{4}}\sum_{k=0}^{4}
\frac{|(2^{2k-1}-1)B_{2k}|}{ k! (4-k)! }
\nonumber \\
&=&-\frac{1}{105 a^{8}},
                                 \label{Dowker}
\end{eqnarray}
where $B_{2k}$ are Bernoulli numbers and $a$ is the radius of the
curvature. It is zero in the optical version of the Nariai metric, as
expected. Moreover, as an additional partial check, we have also
calculated the basic ingredient of the DeWitt method, $[\Box^{5}\sigma
],$ in two different ways, where the biscalar $\sigma(x,x')$ is half
the square of the geodetic distance between points $x$ and $x'.$
Subsequently, making use of the standard formula
\begin{equation}
T^{ab}\,=\,\frac{2}{\sqrt g}\frac{\delta W_{R}}{\delta g_{ab}},
                                  \label{Tdef}
\end{equation}
we have constructed the next-to-leading (i.e. $m^{-4}$) term of the
renormalized stress-energy tensor in a general spacetime. To the best
of our knowledge it is the first attempt to go beyond the first order
(i.e. $m^{-2}$) in the calculations of this type.

The DeWitt method is easily programmable, and the number of terms that
appear at intermediate stages of calculations can be reduced
significantly by a carefully chosen simplification strategy. On the
other hand, the calculations carried out in the  Riemann normal
coordinates are extremely fast~\footnote{It takes a  few minutes to
calculate $[a_{4}]$ and construct the stress-energy tensor using the
covariant the DeWitt method. The analogous calculations in the Riemann
normal coordinates can easily be executed within one minute.}. The
calculations of the coefficient $[a_{4}]$ and its functional
derivatives with respect to the metric  tensor have been carried out
with the aid of FORM~\cite{Vermaseren} and its multithread version
TFORM~\cite{Vermaseren-Tentyukov}. Since the resulting formulas
describing the general second-order stress-energy tensor are lengthy
and rather complex we shall not display them here.

The thus obtained approximate stress-energy tensor can be applied in
any spacetime provided the temporal changes of the geometry are small
and $\lambda_{c}/L\ll 1.$ The effective action approach that we employ
in this paper requires the metric to be positively defined.
Consequently, the stress-energy tensor can be obtained by analytic
continuation of its Euclidean counterpart at the final stage of
calculations.

The first order  (i.e. $m^{-2}$) approximation to the renormalized
stress-energy tensor of the massive scalar, spinor and vector field in
the general spacetime has been constructed in
Refs.~\cite{kocio1,kocio2}. This results generalize the analogous
results obtained earlier by Frolov and Zel'nikov~\cite{FZ1,FZ2,FZ3}
for the vacuum type-D metrics as well as the analytic approximation
obtained by Anderson, Hiscock and Samuel (AHS) for the massive scalar
field in a general static and spherically-symmetric
geometries~\cite{AHS95}. (See also Popov's paper~\cite{Popov}.) The
AHS approximation is equivalent to the Schwinger - DeWitt expansion;
to obtain the lowest (i. e. $m^{-2}$) terms, one has to use sixth-
order WKB expansion of the mode functions.

The range of applicability of such a stress-energy tensor is dictated
by the limitations of the validity of the renormalized effective
action. Numerical calculations reported in Refs.~\cite{AHS95,semiRN}
confirm that the Schwinger-DeWitt method provide a good approximation
of the renormalized stress energy tensor of the massive scalar field
with arbitrary curvature coupling as long as the mass of the field
remains sufficiently large.

The stress-energy tensors constructed in
Refs.~\cite{AHS95,kocio1,kocio2} have been applied in a number of
physically interesting cases, such as various black
holes~\cite{semiRN,kocio1,kocio2,Berej,ja_entr,spin_vect} their
interiors~\cite{interior} and wormholes~\cite{worm}.

In this note we shall calculate the  renormalized stress-energy tensor
of the massive scalar field (in a large mass limit) with an arbitrary
curvature coupling in the geometry of the Schwarzschild black hole up
to $m^{-4}$ terms. We shall also analyze the back reaction problem and
briefly study the quantum corrected Schwarzschild black hole.
Throughout the paper a natural system of units is adopted, although in
some formulas the constants $\hbar,$ $c$ and $G$ have been, for
clarity, restored.

\section{The stress-energy tensor}
Now let us return to Eq.(\ref{Weff}) and retain only the first two
terms. The approximate stress-energy tensor constructed from the
coefficients $[a_{3}]$ and $[a_{4}]$ is, therefore, given by
\begin{equation}
T^{ab} =\frac{1}{32\pi^{2}m^{2}} \frac{2}{\sqrt{g}}\frac{\delta}
{\delta g_{ab}}\int d^{4}x \sqrt{g}[a_{3}]+
\frac{1}{32\pi^{2}m^{4}} \frac{2}{\sqrt{g}}\frac{\delta}{\delta g_{ab}}
\int d^{4}x \sqrt{g}[a_{4}]\equiv
T^{(1)}_{ab} + T^{(2)}_{ab}.
                                        \label{Tog}
\end{equation}
Since the coefficients $[a_{3}]$ and $[a_{4}]$ are, respectively, the
operators of dimension six and eight constructed from the Riemann
tensor, its covariant derivatives up to some prescribed order and
contractions, the result of the functional differentiation of the
effective action with respect to the metric tensor is rather
complicated. Moreover, one expects that any attempt to employ the thus
obtained results for a concrete line element would be,
computationally, a real challenge~\footnote{The total time needed to
calculate components of the stress-energy tensor for a general, static
and spherically-symmetric line element was 15 hours.}. For example,
for a general  static and spherically-symmetric geometry described by
a line element of the form
\begin{equation}
ds^{2} = g_{tt}(r) dt^{2} + g_{rr}(r) dr^{2} + r^{2}\left( d\theta^{2} 
+\sin^{2}\theta d\phi^{2}\right),
                        \label{liniowy}
\end{equation}
the expression describing the next-to-leading term of the stress-
energy tensor, when fully expanded, consists of 2582 primitive terms
for $T^{(2t)}_{t},$  2026 for $T_{r}^{(2)r}$ and $2066$ for
$T_{\theta}^{(2)\theta}.$  This can be contrasted with the number of
the primitive terms in the tensor $T^{(1)b}_{a}:$ 615 for $T^{(1)t}
_{t}$  463 for $T_{r}^{(1)r}$  and 634 $T_{\theta}^{(1)\theta}.$
Fortunately, the final result for a simple metric is, as we shall see,
quite  simple.

Making use of the first-order approximation of the stress-energy
tensor in the Schwarzschild geometry, one easily obtains~\cite{FZ1}
\begin{equation}
T_{t}^{(1)t}\,=\,\frac{M^{2}}{32\pi^{2}m^{2}r^{8}}
\left[\left( 16-{\frac {176\, M}{5\, r}} \right) \eta-{\frac {19}{21}}
+{\frac {626\, M}{315 \,r}}
\right],
                               \label{First_t}
\end{equation}
\begin{equation}
T_{r}^{(1)r}\,=
\,\frac{M^{2}}{32\pi^{2}m^{2}r^{8}}\left[\left( {\frac {48\,M}{5\,r}}
-{\frac {32}{5}} \right) \eta-{\frac {22\,M}{45\,r}}+\frac{1}{3}
\right]
                                \label{First_r}
\end{equation}
and
\begin{equation}
T_{\theta}^{(1)\theta} = T_{\phi}^{(1)\phi} =\frac{M^{2}}{32\pi^{2}m^{2}r^{8}} 
\left[\left( -{\frac {224\,M}{5\,r}}+{\frac {96}{5}}
 \right) \eta+{\frac {734\,M}{315\,r}} -1\right],
                               \label{First_ang}
\end{equation}
where $\eta = \xi-1/6.$ 

Now, let us consider the second term of the equation (\ref{Tog}).
The second-order calculations are, of course, more involved. Fortunately, 
there are massive simplifications for the Ricci-flat geometry and the final 
result in the Schwarzschild geometry is quite simple. Tedious but 
routine calculations give
\begin{equation}
T_{t}^{(2)t} = \frac{M^{2}}{32\pi^{2} m^{4} r^{10}}\left[\left( 144-\frac{752 M}{r} 
+ \frac{6596 M^{2}}{7 r^{2}}\right) \eta
-\frac{44}{5}+\frac{22664 M}{525 r}-\frac{27166 M^{2}}{525 r^{2}}
\right],
                                     \label{Second_t}
\end{equation}
\begin{equation}
T_{r}^{(2)r} = \frac{M^{2}}{32\pi^{2} m^{4} r^{10}}\left[
\left( -\frac {288}{7}-\frac {1164 M^{2}}{7 r^{2}} +\frac {1208 M}{7 r} \right)\eta
+\frac {12}{5}-\frac {776 M}{75 r}+\frac {5506 M^2}{525 r^2}
\right] 
                                     \label{Second_r}     
\end{equation}
and
\begin{equation}
T_{\theta}^{(2)\theta} =T_{\phi}^{(2)\phi} =
\frac{M^{2}}{32\pi^{2} m^{4} r^{10}}\left[\left( \frac {7760 M^{2}}{7 r^{2}} 
- \frac {6084 M}{7 r} + {\frac {1152}{7}} \right) \eta
+ \frac {1304 M}{25 r} - 
\frac {35698 M^{2}}{525 r^{2}}-\frac {48}{5}
\right].
                                      \label{Second_ang}
\end{equation}
The constructed tensor is covariantly conserved, regular and it can 
easily by checked that at the event horizon one has 
$T^{(2)t}_{t}\,=\,T^{(2)r}_{r}.$ 
Moreover, it should be noted that although the general expression 
describing $[a_{4}]$ involves the terms
up to $\xi^{5},$ the final result is linear in $\xi.$
Although, generally speaking, there are no limitations placed on the parameter $\xi ,$
two of its values are particularly appealing, namely $\eta =0$ and $\eta =-1/6$ 
which lead to the conformal and minimal couplings, respectively. 
 \begin{figure}
 \includegraphics[width=11cm]{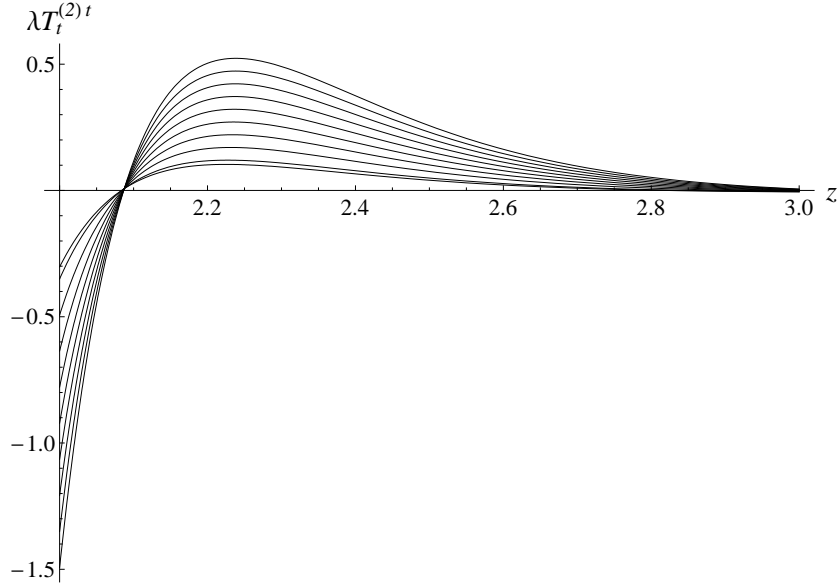}
 \caption{This graph shows the rescaled $T^{(2)t}_{t}$ 
 [$\lambda =(8M)^{4}\pi^{2}m^{4}$] component of the stress-energy tensor of the
 massive scalar field as a function of $z=r/M$ plotted for a few
 exemplary values of the coupling parameter $\xi.$ Top to bottom (at
 the maximum) the curves are plotted for $\xi = 0.2 i$ $(i=0,...,8)$
 and for $\xi =1/6.$
\label{fig1}}
 \end{figure}
 \begin{figure}
 \includegraphics[width=11cm]{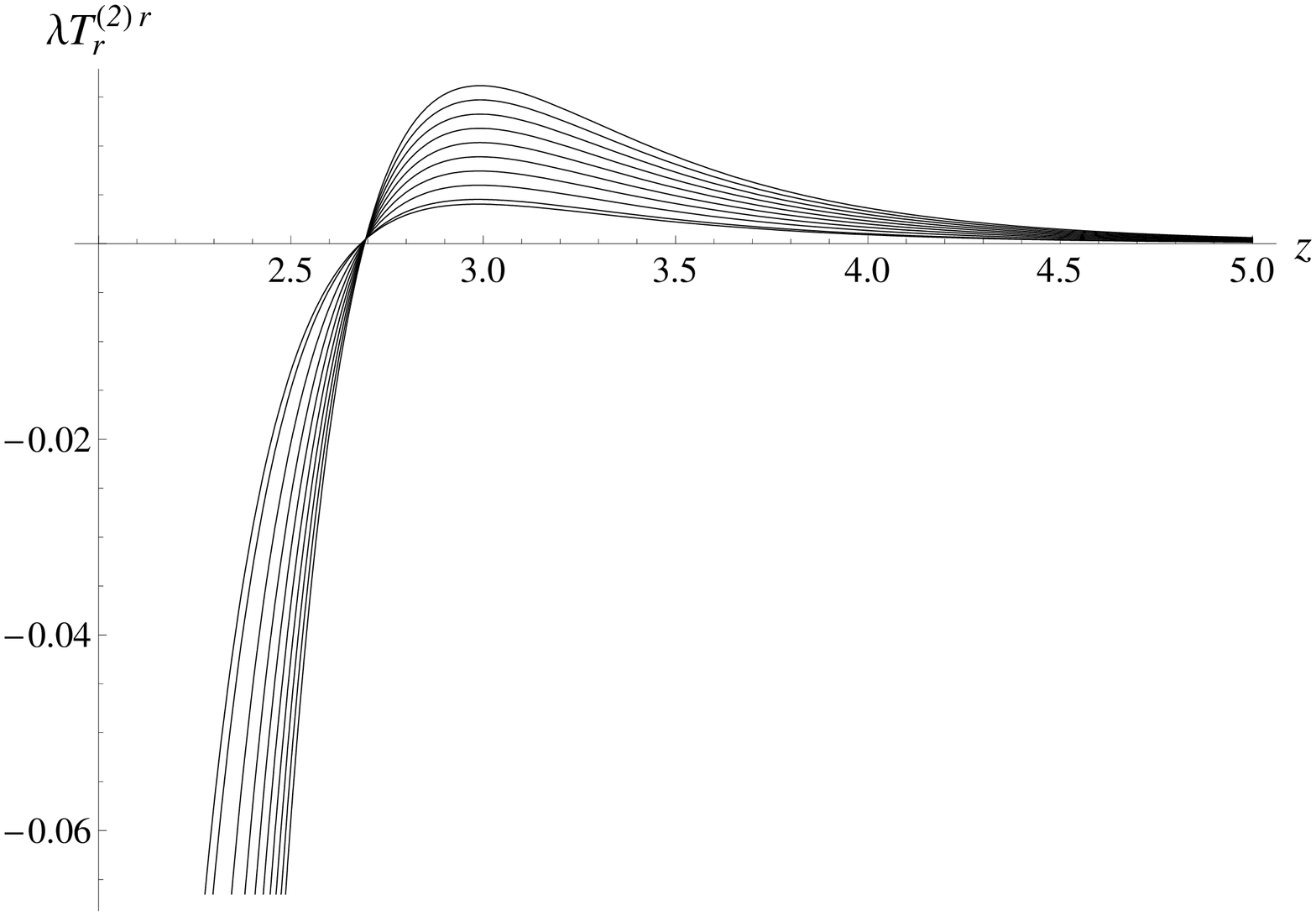}
 \caption{This graph shows the rescaled 
 $T^{(2)r}_{r}$ [$\lambda =(8M)^{4}\pi^{2}m^{4}$]
 component of the stress-energy tensor of the massive 
 scalar field as a function of $z=r/M$ 
 plotted for a few exemplary values of the coupling 
 parameter $\xi.$ Top to bottom 
 (at the maximum) the curves are plotted for $\xi = 0.2 i$ 
 $(i=0,...,8)$ and for $\xi =1/6.$
\label{fig2}}
 \end{figure}
 \begin{figure}
 \includegraphics[width=11cm]{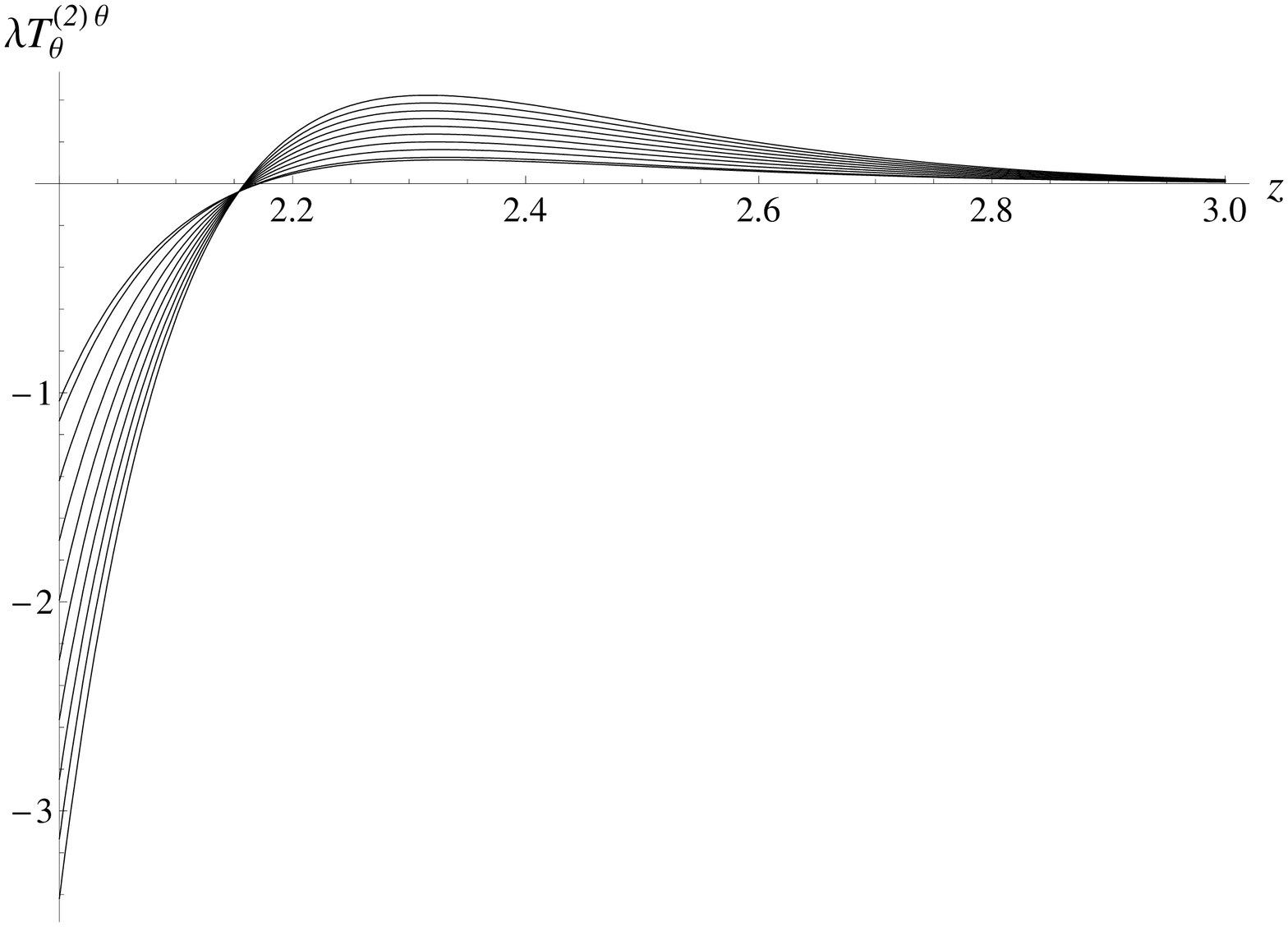}
 \caption{This graph shows the rescaled $T^{(2)\theta}_{\theta}$ 
 [$\lambda =(4M)^{8}\pi^{2}m^{4}$] 
 component of the stress-energy tensor of the massive scalar 
 field as a function of $z=r/M$ 
 plotted for a few exemplary values of the coupling parameter $\xi.$ Top to bottom 
 (at the maximum) the curves are plotted for $\xi = 0.2 i$ $(i=0,...,8)$ 
 and for $\xi =1/6.$
\label{fig3}}
 \end{figure}

In Figs.~1-3 the run of the (rescaled) components of the stress-energy
tensor $T^{(2b)}_{b}$ as functions of $z=r/M$ for a few exemplary
values of the coupling parameter from the range $0\leq \xi \leq 1/6$
is displayed. Although there are strong dependence on $\xi,$ some
general features are common for all the curves. Indeed, the 
$T^{(2)t}_{t}$  is negative at the event horizon and remains so for 
$z \lesssim 2.1 $ and attains a (positive) maximum. Subsequently it 
decreases with
$r$ approaches a (negative) minimum and falls to zero. Inspection of
Fig.~2 shows that  $T^{(2)r}_{r}$ is negative at the event horizon,
approaches a (positive) maximum and fall to zero as $r \to \infty.$
Finally, the run of the angular component (Fig. 3) is qualitatively
similar to that of $T^{(2)t}_{t}.$ The behavior of the stress-energy
for more exotic values of the coupling parameter can easily be
inferred form the general formulas (\ref{Second_t}-\ref{Second_ang}).
Specifically, at the event horizon one has
\begin{equation}
T_{t}^{(2)t} = T_{r}^{(2)r} = \frac{1}{44800\pi^2 m^{4} 
(2M)^8}\left( 1250\eta -53 \right)
\end{equation}
and
\begin{equation}
T_{\theta}^{(2)\theta}  = \frac{1}{26880\pi^2 m^{4} (2M)^8}\left( 1500\eta -109 \right).
\end{equation}

Thus far we have carried out our calculations using the Planck units. 
It is of some interest to restore the constants $c,$ $G$ and $\hbar$ 
in the final expressions describing the renormalized stress-energy 
tensor. Simple manipulations give
\begin{equation}
T_{a}^{(i)b} = A_{(i)} \times f_{(i)a}^{b}(z),
\end{equation}
where $A_{(1)} =G^{2}\hbar^{3}M^{2}/c^{5} m^{2} r^{8},$ 
      $A_{(2)} =G^{2}\hbar^{5}M^{2}/c^{7} m^{4} r^{10},$
and $f_{(i)a}^{b}(z)$ are dimensionless functions of $z=GM/c^{2}r.$

Since the Schwinger-DeWitt approximation is local and the geometry 
at the event horizon is regular, one expects that the stress-energy 
tensor is also regular there. On the other hand, the stress-energy tensor
is regular in the physical sense if it is regular in a coordinate 
system which is well behaved as $r\to r_{+}.$ For example, 
the components of the stress-energy tensor $T_{a}^{b}$  in a 
freely falling frame, denoted here as $T_{(0)(0)}$, $T_{(0)(1)}$ and $
T_{(1)(1)\text{ }\,}$ are 
\begin{equation}
T_{(0)(0)}=\frac{\gamma ^{2}(T_{r}^{r}-T_{t}^{t})}{f}-T_{r}^{r}\text{,}
\label{fram1}
\end{equation}
\begin{equation}
T_{(1)(1)}=\frac{\gamma ^{2}(T_{r}^{r}-T_{t}^{t})}{f}+T_{r}^{r}\text{,}
\end{equation}
\begin{equation}
T_{(0)(1)}=-\frac{\gamma \sqrt{\gamma ^{2}-f}(T_{r}^{r}-T_{t}^{t})}{f}\text{,
}  \label{fram3}
\end{equation}
where $\gamma$ is the energy per unit mass along the geodesic and $f(r) =-g_{tt}(r).$
Inspection of Eqs.~(\ref{fram1}-\ref{fram3})
shows that if  all components of $T_{a}^{b}$ and 
$(T_{r}^{r}-T_{t}^{t})/f$ are finite on the horizon 
the stress-energy tensor in a
freely falling frame is finite as well.	

Now, simple calculations show that the difference between 
radial and time components of the stress-energy
factors
\begin{equation}
T_{t}^{(i)t} -T_{r}^{(i)r} =\left(1-\frac{2M}{r}\right) F^{(i)}(r)
\end{equation}
where
\begin{equation}
F^{(1)}(r) =\frac{M^{2}}{\pi^{2} m^{2} r^{8}}\left(\frac{7}{10}\eta 
- \frac{13}{336} \right),
\end{equation}

\begin{equation}
F^{(2)}(r) =\frac{M^{2}}{\pi^{2} m^{4} r^{10}} 
\left[\left(\frac{81}{14} -\frac{485}{28}\frac{M}{r}
\right)\eta -\frac{7}{20} +\frac{1021}{1050}\frac{M}{r}\right] ,
\end{equation}
$i =1,2,$ and, consequently, both tensors are regular in a physical
sense. Moreover, using our general formula describing the stress-
energy tensor it can be shown that it remains so in any static and
spherically-symmetric spacetime.

\section{The back reaction problem}

 Having constructed the next-to-leading term of the renormalized
stress-energy tensor which depends on a general metric one can analyze
the back reaction of the quantized field upon the  black hole
geometry. It should be emphasized once more that accepting the
approximation (\ref{Tog}) we ignore particle creation which is a
nonlocal effect. To simplify our discussion we shall assume that the
cosmological constant and the renormalized coupling parameters
$\alpha$ and $\beta$ in the quadratic part of the total action
\begin{equation}
S_{q} =\int d^{4} x \sqrt{-g} \left( \alpha C_{abcd}C^{abcd} + \beta R^{2}\right),
\end{equation}
where $C_{abcd}$ is the Weyl tensor, identically vanish. 
The semiclassical Einstein field equations have, therefore, a standard form
\begin{equation}
G_{a}^{b}[g] =8\pi \langle T_{ab}[g]\rangle, 
                      \label{semiE}
\end{equation}
where $\langle T_{ab}[g]\rangle ={\cal O}(\hbar)$ is  
the renormalized stress-energy tensor.

Since the total stress-energy tensor depends functionally on a wide
class of metrics, one can, in principle, construct the self-consistent
solution of the system~(\ref{semiE}). It should be noted however, that
since  the general stress-energy tensor, $T_{a}^{(2)b},$ is
constructed from $[a_{4}]$ it contains the terms up to eight
derivatives of $g_{ab},$ and, consequently, there is a real danger
that the semiclassical equations may lead to physically unacceptable
solutions~\cite{ParkerSimon}. Moreover, the tensor $\langle T_{a}^{b}
\rangle$  is extremely complicated and it is natural that one is
forced to refer to some approximations. Here we shall treat the right
hand side of the semiclassical Einstein field equations as
perturbation. Restricting to the perturbative solutions of the
effective theory may be, therefore, the only one way to obtain the
(approximate) physical solutions.

For the quantized massless fields in the Schwarzschild geometry the
back reaction  program has been initiated by York~\cite{York:1985wp}.
Subsequently, it has been applied in numerous
papers~\cite{Lousto:1988sp,Hochberg:1993xt,Hochberg:1992rd,
Hochberg:1994ps,Anderson:1994hh}, where various aspects of 
the back reaction of the massive fields upon the black hole geometry 
has been studied using the first order approximation to the
stress-energy tensor.

In the semiclassical approach we ignore the effects caused by the
quantized gravitational field simply because they are not known. On
general grounds, however, it is expected that the perturbations of the
classical metrics caused by gravitons should be of the same order as
from the other quantized fields. One can justify neglecting of the
graviton contribution to the total stress-energy tensor taking a large
number of various fields in the calculations. It can be argued that
the contribution of the gravitons would be small as compared with the
total effect caused by other physical fields.

 Now, let us introduce the dimensionless parameter
$\varepsilon$~\cite{Bender} and make the substitution $\langle T_{ab}
[g]\rangle \to \varepsilon \langle T_{ab}[g]\rangle.$ Expanding the
metric tensor as
\begin{equation}
g_{a b}=g^{(0)}_{a b}+ \varepsilon g^{(1)}_{a b} + {\cal O}(\varepsilon^{2}),
                    \label{expandg}
\end{equation}
inserting it into the semiclassical equations (\ref{semiE}) 
and collecting the terms with the like powers of the auxiliary 
parameter, one obtains
\begin{equation}
G_{a}^{b}[g^{(0)}] =0
\end{equation}
and
\begin{equation}
G_{a}^{b}[g^{(1)}] =8 \pi \left( T^{(1)b}_{a}[g^{(0)}] 
+ T^{(2)b}_{a}[g^{(0)}]\right),
\end{equation}
i.e., the modifications of the geometry caused by stress-energy tensor
calculated in the corrected black hole spacetime, $T_{a}^{(1)b}
[g^{(1)}],$ are ignored as these additional terms would be ${\cal O}
(\hbar^{2}).$ In other words we are looking for ${\cal O}(\hbar)$
corrections to the classical solution.

From Eqs.~(\ref{First_t}-\ref{Second_ang}) one sees that the solution
of the back reaction problem reduces to elementary quadratures. First,
let us consider the issue of the integration constants which appear in
solutions of the differential equations. The zeroth-order equations
will yield two integration constants, say, $c_{1}$ and $c_{2},$ which
can be set to $-1$ and $M,$ respectively. Here $M$ is a ``bare" mass
of the black hole and $c_{1}$ can be determined from the condition
$g_{tt}^{(0)} g_{rr}^{(0)}=-1.$ On the other hand, the integration
constant $C_{1}$  appearing in the component $g^{(1)}_{rr}$ of the
metric tensor can be absorbed in a process of the finite
renormalization of mass. Indeed, it can be demonstrated that with the
substitution
\begin{equation}
M = {\cal M}-\frac{1}{2}\varepsilon C_{1} 
                           \label{fin}
\end{equation}
the radial component of the metric tensor can be written as
\begin{equation}
g^{-1}_{rr}(r) =1-\frac{2 \cal{M}}{r} 
+ \frac{{\cal M}^{2}}{5\pi m^{2}r^{6}}\, {\cal P}^{(1)}_{r}(r)
 + \frac{{\cal M}^{2}}{\pi m^{4} r^{8}}\,{\cal P}^{(2)}_{r}(r) 
 + {\cal O} (\varepsilon^{2}),
                              \label{ggrr}
\end{equation}
where ${\cal P}^{(1)}_{r}(r)$ and ${\cal P}^{(2)}_{r}(r)$ 
are given respectively by
\begin{equation}
{\cal P}^{(1)}_{r}(r)=\left({\frac {22{\cal M}}{3 r}} -4 \right) 
\eta-\,\frac {313{\cal M}}{756 r}+\frac {19}{84}
\end{equation}
and
\begin{equation}
{\cal P}^{(2)}_{r}(r)=-\frac {2833 {\cal M}}{2100 r}
+\frac {11}{35}
+\frac {13583 {\cal M}^{2}}{9450 r^{2}}
-
\left( 
\frac {36}{7}
+\frac {1649 {\cal M}^{2}}{63 r^{2}}
-\frac {47 {\cal M}}{2 r}
\right)\eta .
\end{equation}
Similarly, for $g_{tt}$ to ${\cal O}(\hbar)$ one has
\begin{equation}
g_{tt}(r) =-g_{rr}^{-1}(r)\exp\left( 2\varepsilon \psi(r)\right),
\end{equation}
where
\begin{equation}
\psi(r) =\frac{{\cal M}^{2}}{\pi m^{2} r^{6}}\left(\frac{7}{15}\eta-
\frac{13}{504} \right)
+ 
\frac{{\cal M}^{2}}{\pi m^{4} r^{8}}\left[ 
\frac {2042 {\cal M}}{4725 r} -\frac {7}{40}
-\left(\frac {485 {\cal M}}{63 r}
-\frac {81}{28}
\right)\eta\right]+C_{2}.
                                     \label{ps2}
\end{equation}
The integration constant $C_{2}$ can be fixed by demanding that the 
time component of the metric tensor approaches its Minkowskian value 
as $r \to \infty,$ that is equivalent to normalizing the time coordinate 
at infinity.

As before, it is of some interest to restore the physical constants. 
Putting ${\bar{M} =G {\cal M}/c^{2}},$  $l_{Pl} =(\hbar/G c^{3})^{1/2}$ 
and $\lambda_{c} =\hbar/m c$ in Eq.(\ref{ggrr})
the radial component of the metric tensor can schematically be written as
\begin{equation}
g_{rr}^{-1} =1-2 z + \varepsilon \frac{\bar{M}^{2}\lambda_{c}^{2} 
l_{Pl}^{2}}{r^{6}}\, W_{1}(z;\eta)
+ \varepsilon\frac{\bar{M}^{2}\lambda_{c}^{4}l_{Pl}^2}{r^{8}}\, W_{2}(z;\eta),
                              \label{jedn}
\end{equation}
where $W_{1}$ and $W_{2}$ are simple polynomials depending parametrically on $\eta$ 
and their exact form can easily be inferred from Eq.~(\ref{jedn}).
A similar expression can  be constructed for $g_{tt},$ and the result 
can be schematically
 written in the form
\begin{equation}
g_{tt} =-1+ 2 z + \varepsilon \frac{\bar{M}^{2}\lambda_{c}^{2} 
l_{Pl}^{2}}{r^{6}}\, V_{1}(z;\eta)
+ \varepsilon\frac{\bar{M}^{2}\lambda_{c}^{4}l_{Pl}^2}{r^{8}}\, V_{2}(z;\eta),
                              \label{jedn_2}
\end{equation}
where $V_{i}$ comprise another pair of simple polynomials.

The location of the event horizon of the quantum-corrected 
Schwarzschild black hole 
is determined by the equation $g_{tt}(r_{+})=0.$ 
Putting $r_{+} =r^{(0)}_{+} + \varepsilon r^{(1)}_{+}$
one concludes that $r_{+}$  is given by
\begin{equation}
r_{+} = 2 \bar{M} +\frac{\lambda_{c}^{2}l_{Pl}^{2}}{480\pi 
\bar{M}^{3}}\left(\eta -\frac{29}{504} \right)
+ \frac{\lambda_{c}^{4}l_{Pl}^{2}}{2016\pi \bar{M}^{5}}
\left(\frac{17}{1200} -\eta\right).
                 \label{reh}
\end{equation}
The Euclidean version of the line element (\ref{liniowy}) obtained 
with the aid of the Wick rotation has no conical singularity provided 
the ``time" coordinate is periodic with a period $\beta$ given  by
\begin{equation}
\beta =\lim_{r \to r_{+}} 4\pi \left(g_{tt} g_{rr} \right)^{1/2} 
\left( \frac{d}{dr}g_{tt}\right)^{-1}.
\end{equation}
For the quantum-corrected metric (\ref{ggrr}-\ref{ps2}), the period is
\begin{equation}
\beta =8\pi {\cal M} -\frac{\varepsilon}{30240 {\cal M}^{3} m^{2}} 
-\frac{11\varepsilon}{151200 {\cal M}^{5} m^{4}}
\end{equation}
to the first order in $\varepsilon.$ 

The surface gravity, $\kappa,$ which is proportional to the temperature 
of the black hole can be calculated (for the Lorentzian metric) 
from a simple relation
\begin{eqnarray}\kappa^{2} &=& {-\frac{1}{2} k_{a ;b} k^{a ; b}}_{|r=r_{+}}\nonumber\\
           &=& -{\frac{1}{4}\left(g_{tt}g_{rr}\right)^{-1} 
           \left( \frac{d g_{tt}}{dr} \right)^{2}}_{|r=r_{+}},
\end{eqnarray}
where $k^{a}$ is a timelike Killing vector. 
The Hawking temperature,  $T_{H}=\kappa/2\pi ,$ 
is, therefore, given by
\begin{eqnarray}
T_{H} &=& \frac{1}{4\pi}\left(-g_{tt} g_{rr}\right)^{-1/2}
\lvert\frac{d g_{tt}}{dr}\rvert_{|r=r_{+}} =\frac{1}{\beta} \nonumber \\
&=&\frac{1}{8\pi {\cal M}} +\frac{\varepsilon}{\pi^{2}m^{2}
(4{\cal M})^{5}}\left(\frac{1}{1980}+\frac{11}{9450{\cal M}^{2} m^{2}}\right).
\end{eqnarray}
It should be noted that the temperature, $T_{H},$ when expressed in terms
of the total mass of the system  as seen by a distant observer, is independent of
the coupling constant.

On the other hand, one can express the results (\ref{ggrr}-\ref{ps2}) 
in terms of the horizon defined mass, $M_{H} =r_{+}/2$ which, of course,
differs from the total mass of the system as seen by a distant observer.
It can be achieved, for example, by inverting Eq.~(\ref{reh}) and  
the elementary manipulations give 
\begin{equation}
{\cal M} =M_{H} -\frac{\varepsilon}{960\pi m^{2} M_{H}^{3}}
\left(\eta -\frac{29}{504} \right)-\frac{\varepsilon}
{4032\pi m^{4} M_{H}^{5}}\left(\frac{17}{1200} -\eta\right).
                                 \label{zam}
\end{equation}
Consequently, one can systematically substitute 
Eq.~(\ref{zam}) in Eqs.(\ref{ggrr}-\ref{ps2}), 
expand and finally linearize the thus obtained results. 

Equally well, one can start with 
a slightly different representation of the line element putting
\begin{equation}
g_{tt}(r) = -\exp (2\psi(r))\left(1-\frac{2 M(r)}{r}\right), \,\,\,\,\,
g_{rr} =1-\frac{2 M(r)}{r} 
                              \label{nowy}
\end{equation}
expanding the functions $M(r)$ and $\psi(r)$ into the power series
\begin{equation}
M(r) = \sum_{i=0}^{k} M_{i}(r) \varepsilon^i
\end{equation}
\begin{equation}
\psi(r) = \sum_{i=1}^{k} \psi_{i}\varepsilon^{i}
\end{equation}
and retaining only the linear terms. 
Now, accepting the boundary condition $\psi_{1}(\infty) =0,$ repeating the calculations 
for the line element (\ref{nowy}),
with the integration constants appearing in the solution 
for $M(r)$ determined either from
$M_{0}(\infty) ={\cal M}$ and $M_{1}(\infty)=0$
or from  $M_{0}(r_{+})=r_{+}/2$ and $M_{1}(r_{+})=0,$  
one can easily reconstruct all our previous results.

\section{final remarks} 
In this paper we report our calculations of the next-to-leading term
of the renormalized stress-energy tensor of the quantized massive
scalar field in a large mass limit. To achieve this, we have
calculated the effective action constructed form the (integrated)
coincidence limit of the coefficients $a_{3}(x,x')$ and $a_{4}(x,x'),$
and, subsequently, we have calculated the approximate stress-energy
tensor by functional differentiation of the thus obtained action with
respect to the metric tensor. The obtained stress-energy tensor can be
employed in any spacetime provided the condition $\lambda_{C}/L \ll 1
$ holds. The general formulas describing the stress-energy tensor are
extremely complex, but, when applied to the Schwarzschild geometry,
they yield remarkably simple result, which is the main result of this
paper. The general stress-energy tensor has been used in the analysis
of the back reaction of the quantized field upon the geometry of the
Schwarzschild black hole.

Finally, we indicate a few possible  directions of investigations. 
First, it would be interesting to examine the vacuum polarization effects
in more complex backgrounds, as for example, the spacetime of the 
electrically charged black holes with or without the cosmological constant.
Further, the numeric approach to the back reaction would certainly
strengthen our understanding of the problem. 
This group of problems is
actively investigated and the results will be published elsewhere.


\end{document}